\documentstyle[psfig]{cupconf}

\ifoldfss
\else
  \ifnfssone
    \newmathalphabet{\mathit}
      \addtoversion{normal}{\mathit}{cmr}{m}{it}
      \addtoversion{bold}{\mathit}{cmr}{bx}{it}
    \newmathalphabet{\mathcal}
      \addtoversion{normal}{\mathcal}{cmsy}{m}{n}
    \else
    \ifnfsstwo
    \fi
  \fi
\fi

\def\hexnumber#1{\ifcase#1 0\or1\or2\or3\or4\or5\or6\or7\or8\or9\or
 A\or B\or C\or D\or E\or F\fi }

\makeatletter
\ifx\CUP@mtlplain@loaded\undefined
\else
\fi
\makeatother
 \makeatletter
 \ifx\CUP@mtlplain@loaded\undefined
   \font\tenbmi=cmmib10 at 10pt
   \font\sevenbmi=cmmib10 at 7pt
   \font\fivebmi=cmmib10 at 5pt

   \newfam\bmifam
   \textfont\bmifam=\tenbmi
   \scriptfont\bmifam=\sevenbmi
   \scriptscriptfont\bmifam=\fivebmi
   
 \fi
 \makeatother
\ifnfsstwo

\fi
\ifnfssone

\fi
\ifoldfss

\fi

\mathchardef\varLambda="0103

\makeatletter
\ifx\CUP@mtlplain@loaded\undefined
\else
\fi
\makeatother
\makeatletter
\ifx\CUP@mtlplain@loaded\undefined
  \font\tenbms=cmbsy10
  \font\sevenbms=cmbsy10 at 7pt
  \font\fivebms=cmbsy10 at 5pt
  \newfam\bmsfam
  \textfont\bmsfam=\tenbms
  \scriptfont\bmsfam=\sevenbms
  \scriptscriptfont\bmsfam=\fivebms

  \edef\bsy@{\hexnumber\bmsfam}
  \mathchardef\bnabla="0\bsy@72
\fi
\makeatother
\title[ Deep Photometric Survey of IC2391]{ Deep Photometric Survey of IC2391}

\author[D. Barrado y Navascu\'es {\it et al.\/}]%
{D.\ns B\ls A\ls  R\ls R\ls A\ls D\ls O\ns y\ns
N\ls A\ls V\ls A\ls S\ls C\ls U\ls \'E\ls S$^1$\thanks{Present address: 
MPI f\"ur Astronomie, K\"onigstuhl 17, D-69117 Heidelberg, Germany},\ns
J.\ns R.\ns S\ls T\ls A\ls U\ls F\ls F\ls E\ls R\ns$^1$\\
\and \ns  C.\ns B\ls R\ls I\ls C\ls E\ls \~N\ls O$^2$}

\affiliation{$^1$Harvard-Smithsonian Center for Astrophysics, 60 Garden St.,
 Cambridge,  MA02138, USA\\[\affilskip]
$^2$Yale University, USA}

\begin{document}
\ifnfssone
\else
  \ifnfsstwo
  \else
    \ifoldfss
      \let\mathcal\cal
      \let\mathrm\rm
      \let\mathsf\sf
    \fi
  \fi
\fi

\maketitle

\begin{abstract}

We present a deep survey of the young southern cluster IC2391. We have
covered 2 sq. degrees in the R,I filter. We have selected several 
dozens VLM stars and BD  candidates based on the position of these objects
in the Color-Magnitude Diagram.

\end{abstract}

\firstsection % if your document starts with a section,
\section{Introduction}

The open cluster IC2391 ($\alpha$=8h 38.8m, $\delta$=-52$^\circ$53')
has adequate characteristics to carry out a survey in order to look for
very low mass stars (VLM) and brown dwarfs (BD) candidates. It is 
a very young cluster, considerably younger than the Pleiades
(the standard age is $\sim$30 Myr, versus  the often quoted standard age
for the Pleiades of 70-80 Myr). In addition, it distance is approximately
the same as the Pleiades and it is not spread over a large area, making 
feasible to cover a significant fraction of the cluster in few nights.

\begin{figure} 
  \vspace{-3cm}
  \centerline{\hspace{-2.5cm}\psfig{figure=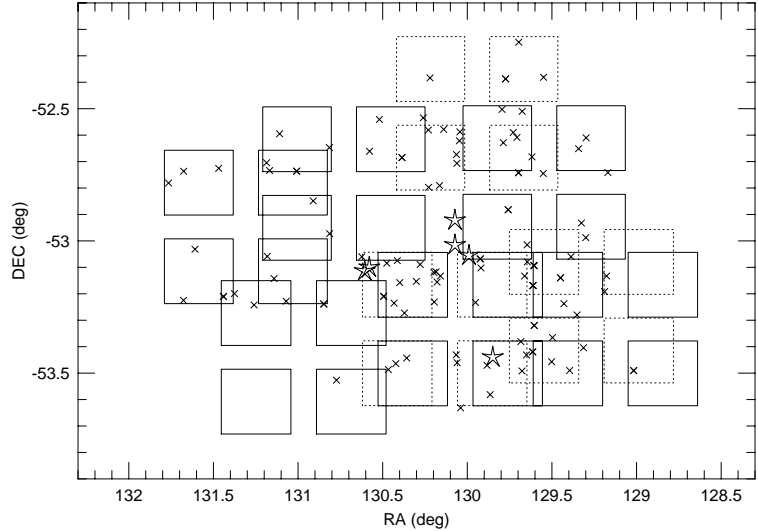,width=9.0cm}}
  \caption{Graphic representation of 
 the area of the sky around IC2391. The six stars brighter than V=6$^{\rm mag}$
are shown. Our fields appear as solid (\#1,\#2,\#4,\#5,\#6) and dotted
(\#1',\#3',\#5') squares. Candidates are shown as crosses.} 
\end{figure} 

\begin{table}
\begin{center}
\begin{tabular}{lccrrcrr}
\hline
Field  &  RA   &   DEC &\multicolumn{2}{c}{Exp. Time} & & \multicolumn{2}{c}{Mag. limit} \\
\cline{4-5}  \cline{7-8}
       & \multicolumn{2}{c}{(2000.0)} & R$_c$  & I$_c$ && R$_c$  & I$_c$  \\
\cline{2-3}
       &(h m s)&($^\circ$ ' '') &   (sec)  &  (sec) && & \\
\hline
$\#$1     &08:40:10&-53:20:01   &   2880   & 1200 && 23.1 & 21.7 \\
$\#$1'    &08:40:32&-53:19:58   &   1200   & 600  && 23.1 & 21.5\\
$\#$2     &08:43:52&-53:26:25   &   1440   & 600  && 23.1 & 21.3\\
$\#$3     &08:38:11&-52:46:44   &   960    & 960  && 22.8 & 19.4\\
$\#$3'    &08:39:46&-52:31:03   &   1200   & 600  && 23.5 & 22.0\\
$\#$4     &08:42:55&-52:46:59   &   600    & 1440 && 22.8 & 20.8\\
$\#$5     &08:36:30&-53:20:00   &   1200   & 600  && 23.1 & 21.8\\
$\#$5'    &08:37:04&-53:14:49   &   274    & 400  && 23.0 & 21.8\\
$\#$6     &08:45:14&-52:56:50   &   1200   & 600  && 23.5 & 22.1\\
\hline
\end{tabular}
  \caption{Observed fields}
\end{center}
\end{table}

We  carried out a deep photometric survey in January  1998, 
using  the cousin R and I filters, and covering a large 
fraction of the cluster, around 2 square degrees. We used the 
CTIO 4 meter Blanco telescope and the Big Throughput Camera, a mosaic
detector composed by 4 different CCDs of 2048$\times$2048 pixels each.
One CCD covers  14.7$\times$14.7 arc-minutes and their edges are
separated from each other by a distance of 5.4 arc-minutes.

We observed 9 different fields around the center of the cluster, avoiding
the bright stars in the center. Figure 1 shows the portions of the sky
 covered by our survey, together the 6 brightest stars on the field.
Table 1 lists the positions of the center of each field. Exposure times
are also listed in the same table, as well as the magnitude limits.

Since our nights were not photometric, we used an observation collected 
several days before at the CTIO 0.9m telescope to calibrate the data.
Additional calibrations were derived from other observations taken 
on April. All agree quite well and we have estimated the error
in our photometry as better than 0.05 mag.

\begin{figure} 
  \vspace{-3cm}
  \centerline{\hspace{-2.5cm}\psfig{figure=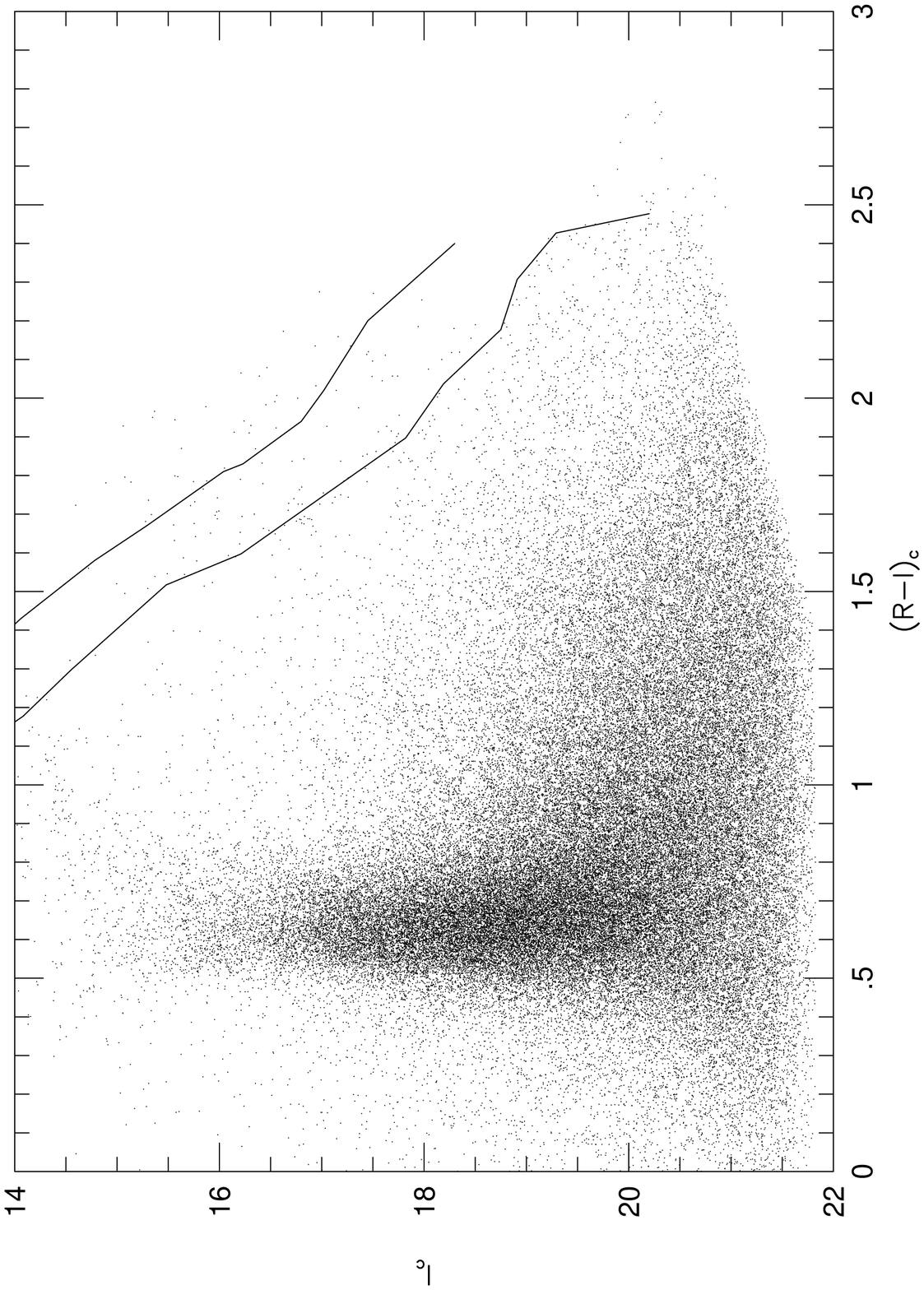,width=9.0cm}}
  \caption{ Color-Magnitude diagram. Our member candidates are
shown as filled circles. A ZAMS (empirical, obtained from field stars)
 and a 30 Myr isochrone
 (D'Antona \& Mazzitelli 1994) appear as solid lines.
We used the 30 Myr isochrone to select the candidates. Note the gap between
18 $\le$ I$_c$ $\le$ 19.5 magnitudes.
} 
\end{figure} 

\section{ Membership in the Cluster}

Candidates to be members of the clusters were selected based on the 
R,I Color-Magnitude diagram (CMD). Figure 2 shows the objects detected in our
survey (dots). An empirical Zero Age Main Sequence (ZAMS) and
 a 30 Myr isochrone (D'Antona \& Mazzitelli 1994) are also  displayed.
The ZAMS was built by  tracing the lower envelope of Color-Magnitude Diagram 
of nearby dK and dM stars (Leggett 1992).
 We have used these curves to select our best candidate members. Since the
real age of the cluster is not known (Stauffer et al. 1998 have obtained an
age of 125 Myr for the Pleiades based on the lithium depletion boundary. See
Barrado y Navascu\'es et al. 1999 for a discussion about the new age scale for
young open clusters) and, therefore, where the MS lies, we selected stars 
below and above the 30 Myr isochrone, but considerably brighter than the ZAMS.
 Based on the comparison
with the data from Simon \& Patten (1996) -and references therein-
 and Pavlovsky \& Patten (1998), represented as filled and empty
triangles in Figure 3a, respectively,  our best candidates for
IC2391 membership are those VLM stars and BD which are located in the upper
part of our CMD. However, due to the lack of additional information
(e.g. proper motions, spectroscopy, location of the lithium depletion 
boundary), we cannot rule out that, for the same color,  fainter objects 
in our sample of candidates 
belong to the cluster. A spectroscopic follow-up of this sample
will allow us to establish the locus of the MS of the cluster and an
accurate age.

A comparison between our candidates and members of the clusters
$\alpha$ Per and the Pleiades is presented in Figure 3b. We used
the distances and reddening by Pinsonneault et al. (1998), derived 
from fitting a MS to Color-Magnitude diagrams of actual members.
We have plotted the data using other values of the distances
[as measured with Hipparcos -Mermilliod et al. (1997)- and from 
Lyng\aa{ } (1987)] and the figure does not change in a significant way. 
If our survey contains an important fraction of IC2391 members and 
if our assumed distances and reddenings are correct, then IC2391
is not  apparently significantly younger than $\alpha$ Per.
Recent measurement of lithium in $\alpha$ Per members
(Barrado y Navascu\'es et al. 1999)
indicates that it could be as old as $\sim$85 Myr old.

\begin{figure} 
% \vspace{16.5pc}
 \vspace{-1cm}
  \centerline{\hspace{-1.8cm}\psfig{figure=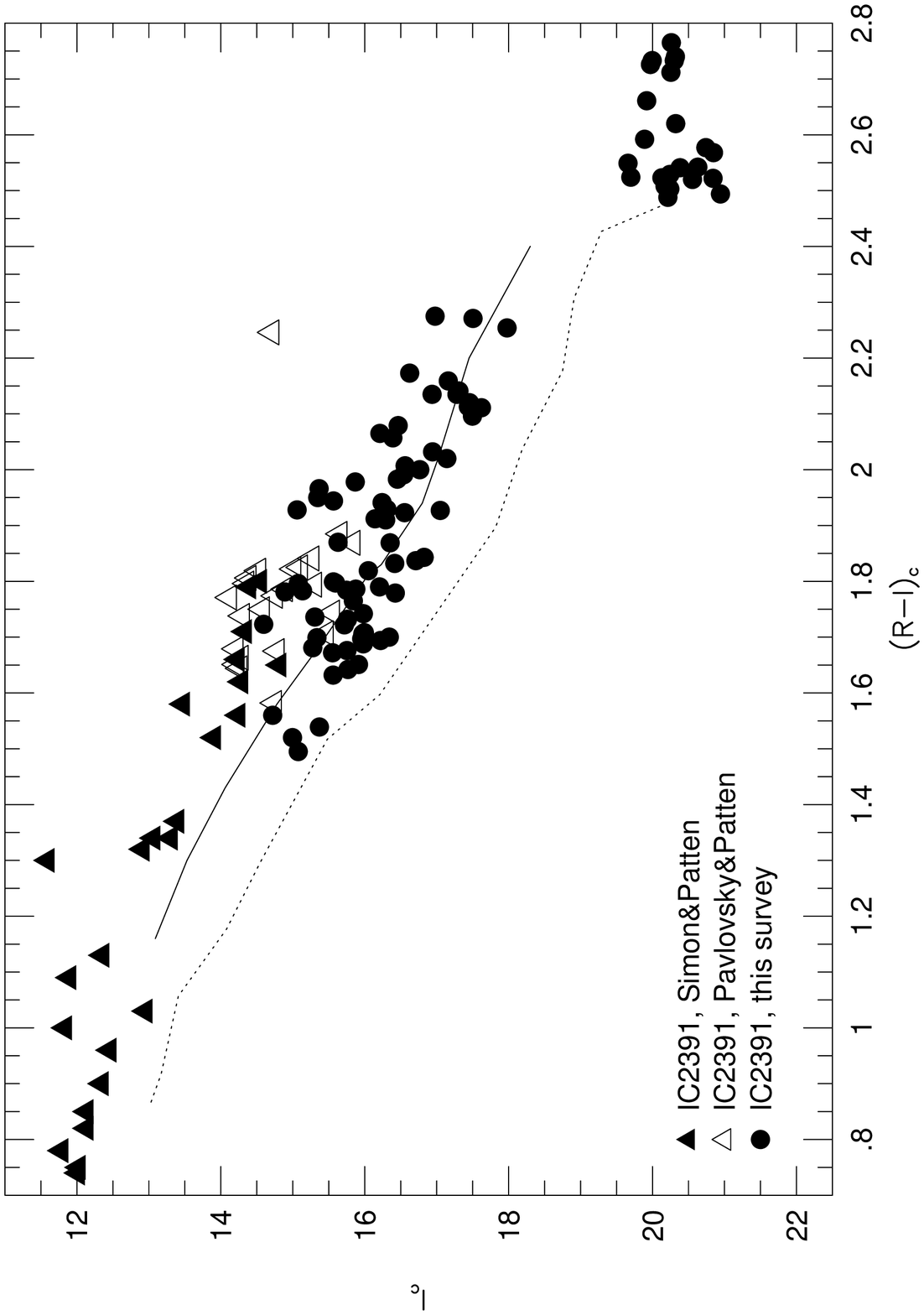,width=6.0cm}
\hspace{8mm}\psfig{figure=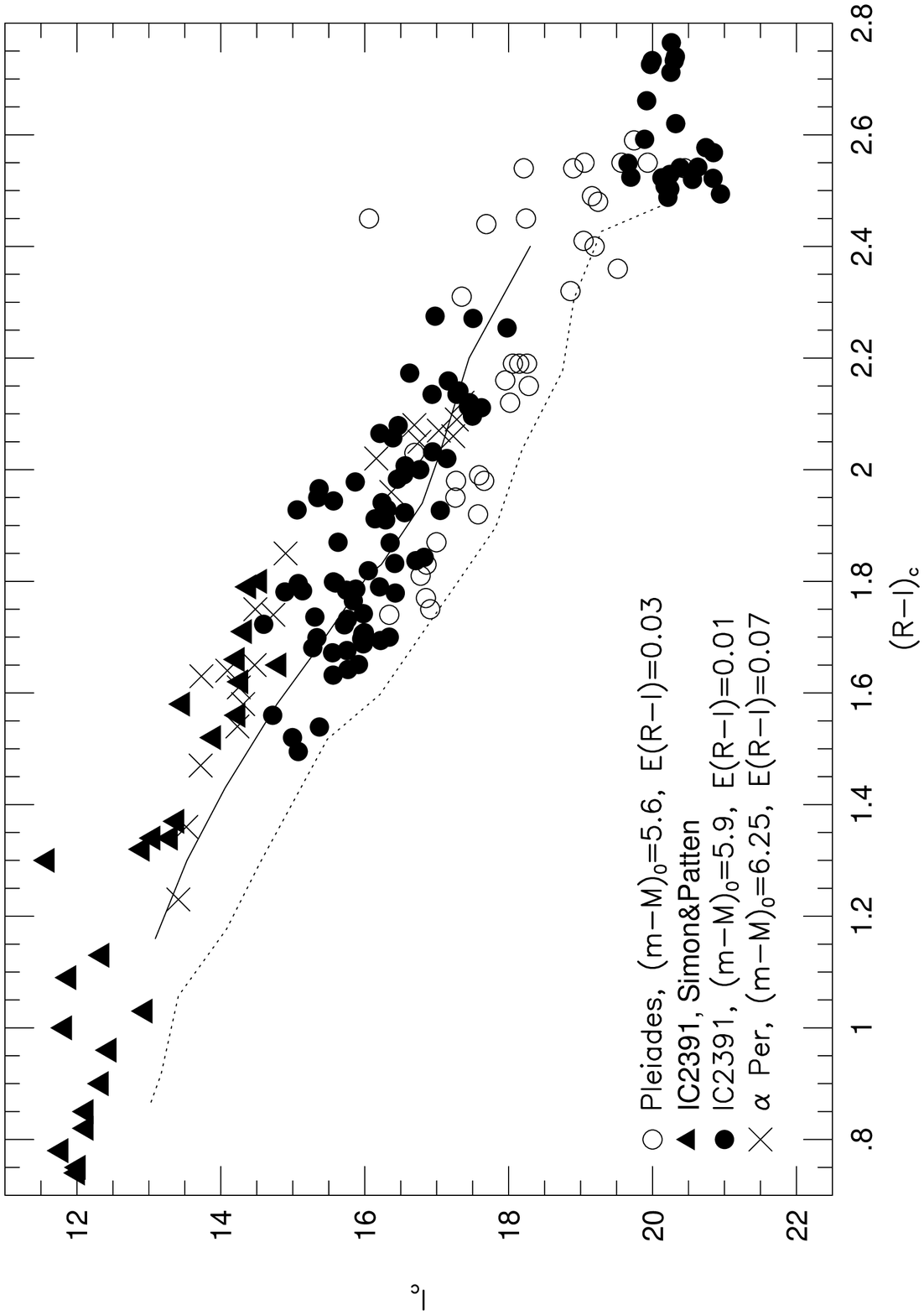,width=6.0cm} }
  \caption{
{\bf a}   Comparison between IC2391 candidates
 by Simon \& Patten
(filled triangle), Pavlovsky \& Patten candidates (empty triangles), 
and our IC2391 VLM and BD candidates (filled
circles). 
{\bf b}  Comparison between Pleiades members (empty circles),
$\alpha$ Per members (crosses),  and our IC2391 VLM and BD candidates (filled
circles). These clusters have standard ages of $\sim$70, 50 and 30 Myr, 
respectively.
} 
\end{figure} 

The lack of objects in the IC2391 locus in the range
 18 $\le$ I$_c$ $\le$ 19.5 is puzzling. It could indicate a sudden
cut--off in the mass function of the clusters. Using the D`Antona
\& Mazzitelli (1994) models, it would appear for BD at 0.03 M$_\odot$.
It is also intriguing the detection of a significant number of objects
redder than the ZAMS and fainter than I$\sim$19.5 mag. These are real 
detections and the photometry looks good enough. We have tried to perform
several test to confirm or reject their membership.
Firstable, there is a clear gap between the field stars
and those objects detected in the position of the IC2391 locus.
In addition, our candidates are concentrated at the center of the cluster.
 Moreover, despite the scarcity of data,
 the gap does not appear for field stars slightly blueward of the ZAMS.
An histogram for stars in field \#1 having 
19.5 $\le$ I$_c$ $\le$ 20.5 mag. shows that there is a clear
 drop at (R--I)$\sim$2.2 and that there is a 
high number of objects in the (R-I) bins corresponding to  the position
of IC2391. This situation induces us to consider them as possible extremely
 low mass BD, members of the  cluster,  
which deserve further observations to try to confirm their status.

\begin{acknowledgments}
DBN acknowledge the support by 
the ``Real Colegio Complutense at Harvard University'', the
MEC/Fulbright Commission, and the European Union.
\end{acknowledgments}

\end{document}